\documentclass[a4paper,11pt]{amsart}
\begin{document}
\hyphenation{gra-vi-ta-tio-nal re-la-ti-vi-ty va-cuum
Schwarz-schild re-fe-ren-ce re-la-ti-ve sin-gu-la-rity
gra-vi-ta-tion}
\title[On Serini's relativistic theorem] {{\bf  On Serini's relativistic theorem}}
\author[Angelo Loinger]{Angelo Loinger}
\thanks{To be
published on \emph{Spacetime \& Substance.} \\email:
angelo.loinger@mi.infn.it\\ Dipartimento di Fisica, Universit\`a
di Milano, Via Celoria, 16 - 20133 Milano (Italy)}

\begin{abstract}
I expound here in a more detailed way a proof of an important
Serini's theorem, which I have already sketched in a previous
Note. Two related questions are briefly discussed.
\end{abstract}

\maketitle


\vskip1.20cm
\noindent \textbf{1}. - \textbf{Introduction}
\par  In the Note ``On the beams of wavy metric tensors'' \cite{1}
I sketched a simple and concise proof of an important Serini's
theorem (1918). Serini's original demonstration was subsequently
generalized by Einstein and Pauli (1943) and by Lichnerowicz
(1946) (see \cite{2}). Independently, a different proof was given
by Fock \cite{2bis}.

\par I wish now to expound again, in a more detailed way, my proof
of the above theorem, and to solve an apparent contradiction with
the main result of the paper ``Regular solutions of Schwarzschild
problem'' \cite{3}. The Appendices contain some remarks concerning
the so-called gravitational waves.

\vskip0.50cm
\noindent \textbf{2}. - \textbf{The theorem} \par Serini's theorem
affirms the non-existence of \emph{regular} (i.e. without
singularities) time independent solutions of Einstein field
equations for the perfectly empty space, $R_{jk}=0$,
$(j,k=0,1,2,3)$, that become pseudo-Euclidean at spatial infinity.
The unique time independent solution of $R_{jk}=0$ is the trivial
solution $g_{jk}=\textrm{const}$. Thus, in the time independent
case, $R_{jk}=0$ imply $R_{jklm}=0$, the vanishing of Riemann
curvature tensor of the spacetime manifold.

\vskip0.50cm
\noindent \textbf{3}. - \textbf{Proof of the theorem} \par
\emph{a}) As it was remarked by Hilbert \cite{4}, we can always
choose a Gaussian normal (or synchronous \cite{5}) system of
coordinates for the solution of \emph{any} relativistic problem.
In their treatise \cite{5} Landau and Lifchitz explain in a
detailed way the interesting properties of this reference system.
I shall follow their treatment, but with some slight differences
of notations.

\par A Gaussian normal (or synchronous) reference frame can be
defined by the conditions:

\begin{equation} \label{eq:one}
    g_{00}=1, \qquad g_{0\alpha}=0, \quad (\alpha=1,2,3);
\end{equation}

accordingly:

\begin{equation} \label{eq:two}
    \textrm{d}s^{2}= (\textrm{d}x^{0})^{2} +
    g_{\alpha\beta}(\textbf{x},x^{0})
    \textrm{d}x^{\alpha}\textrm{d}x^{\beta};
\end{equation}

putting $g_{\alpha\beta}\equiv-h_{\alpha\beta}$, we have:

\begin{equation} \label{eq:twoprime}
    \textrm{d}s^{2}= (\textrm{d}x^{0})^{2} -
    h_{\alpha\beta}(\textbf{x},x^{0})
    \textrm{d}x^{\alpha}\textrm{d}x^{\beta}.
    \tag{$2'$}
\end{equation}

It is  easy to see that the time lines \emph{coincide} with the
spacetime geodesics. Henceforth, all the operations of index
displacement and covariant derivative concern \emph{only} the
three-dimensional space with the metric tensor $h_{\alpha\beta}$.
If

\begin{equation} \label{eq:three}
    \kappa_{\alpha\beta} :=
    \frac{\partial h_{\alpha\beta}}{\partial x^{0}},
\end{equation}

the components of the Ricci-Einstein tensor $R_{lm}$,
$(l,m=0,1,2,3),$ are:

\begin{equation} \label{eq:four}
    R_{00}= -\frac{1}{2} \frac{\partial \kappa_{\alpha}^{\alpha}}{\partial x^{0}}
    -\frac{1}{4} \kappa_{\alpha}^{\beta} \kappa_{\beta}^{\alpha} ,
\end{equation}

\begin{equation} \label{eq:fourprime}
    R_{0\alpha}= \frac{1}{2} \left(\kappa_{\alpha;\beta}^{\beta} -\kappa_{\beta;\alpha}^{\beta}\right) ,
    \tag{$4'$}
\end{equation}

\begin{equation} \label{eq:foursecond}
    R_{\alpha\beta}= \frac{1}{2} \frac{\partial \kappa_{\alpha\beta}}{\partial x^{0}}
    +\frac{1}{4} \left(\kappa_{\alpha\beta} \kappa_{\gamma}^{\gamma} - 2\kappa_{\alpha}^{\gamma} \kappa_{\beta\gamma}\right)
     + \textrm{P}_{\alpha\beta} ,
    \quad \; \tag{$4''$}
\end{equation}

where $\textrm{P}_{\alpha\beta}$ is the three-dimensional analogue
of $R_{lm}$.

\par The Riemann curvature tensor $R_{lmrs}$ is given by:

\begin{equation} \label{eq:five}
    R_{\alpha\beta\gamma\delta}= \textrm{P}_{\alpha\beta\gamma\delta}
    +\frac{1}{4} \left(\kappa_{\alpha\gamma} \kappa_{\beta\delta} - \kappa_{\alpha\beta} \kappa_{\gamma\delta}\right) ,
\end{equation}

\begin{equation} \label{eq:fiveprime}
    R_{0\alpha\beta\gamma}= \frac{1}{2} \left(\kappa_{\alpha\beta;\gamma} - \kappa_{\alpha\gamma;\beta}\right) ,
     \tag{$5'$}
\end{equation}

\begin{equation} \label{eq:fivesecond}
    R_{0\alpha0\beta}= -\frac{1}{2} \frac{\partial \kappa_{\alpha\beta}}{\partial x^{0}}
    + \frac{1}{4} \kappa_{\alpha\gamma} \kappa_{\beta}^{r} ,
     \tag{$5''$}
\end{equation}

where $\textrm{P}_{\alpha\beta\gamma\delta}$ is the
three-dimensional analogue of $R_{lmrs}$.

\vskip0.10cm
\par \emph{b}) For a \emph{time-independent} metric
tensor $h_{\alpha\beta}(\textbf{x})$, we have:

\begin{equation} \label{eq:six}
    R_{00}= R_{0\alpha}=0 ,
\end{equation}

\begin{equation} \label{eq:sixprime}
    R_{\alpha\beta}= \textrm{P}_{\alpha\beta} ;
     \tag{$6'$}
\end{equation}

\begin{equation} \label{eq:seven}
    R_{\alpha\beta\gamma\delta}= \textrm{P}_{\alpha\beta\gamma\delta} ,
\end{equation}

\begin{equation} \label{eq:sevenprime}
    R_{0\alpha\beta\gamma}= R_{0\alpha0\beta}=0 ;
     \tag{$7'$}
\end{equation}

now, it is (see e.g. Fock \cite{2bis}, App. G):

\begin{equation} \label{eq:eight}
    \textrm{P}_{\alpha\beta\gamma\delta}= \left(\textrm{P}^{\rho\sigma} -\frac{1}{2}
    h^{\rho\sigma}\textrm{P}\right)
    E_{\rho\alpha\beta} E_{\sigma\gamma\delta},
\end{equation}

where

\begin{equation} \label{eq:eightprime}
    E_{\alpha\beta\gamma}:= h^{1/2}e_{\alpha\beta\gamma} ,
     \tag{$8'$}
\end{equation}

if $h\equiv \det\|h_{\alpha\beta}\|$, and $e_{\alpha\beta\gamma}$
is a system of antisymmetric quantities with $e_{123}=1$.

\vskip0.10cm
\par \emph{c}) For a perfectly empty space, we have:

\begin{equation} \label{eq:nine}
    R_{lm}= 0,
\end{equation}

and therefore, as an immediate consequence of eqs. (\ref{eq:six}),
$(6')$, (\ref{eq:seven}), $(7')$, (\ref{eq:eight}), $(8')$:

\begin{equation} \label{eq:ten}
    R_{lmrs}= 0, \qquad \textrm{\emph{q.e.d.}} ;
\end{equation}

the unique \emph{time-independent} solution of $R_{lm}=0$ is
$g_{lm}=\textrm{const.}$ This result is obviously quite intuitive,
because the curvature of spacetime is created by matter, and if
the matter is absent \ldots  --

\par (Remark that the above proof does not require the hypothesis
that $g_{jk}$ is pseudo-Euclidean at spatial infinity.)

\vskip0.50cm
\noindent \textbf{4}. - \textbf{An apparent contradiction}
\par At the first sight, it seems that Serini's theorem denies, in
particular, the existence of those \emph{regular} solutions of
Schwarzschild problem -- i.e., of the problem to determine the
gravitational field of a point mass at rest -- which have been
exhibited in paper \cite{3}. However, the contradiction is only
apparent: indeed, \emph{all} forms of solution of Schwarzschild
problem are in reality relative to a matter tensor $T_{jk}$
different from zero, and precisely: to a matter tensor involving a
Dirac delta-distribution \cite{6}, or to the matter tensor of the
limiting case of a concentrated mass, according to Fock's
procedure \cite{7}, which was also followed in \cite{3}.

\vskip0.70cm
\begin{center}
\noindent \emph{\textbf{APPENDIX}} \textbf{A}
\end{center}

In sect.\textbf{3}. of paper \cite{1} I have given an
\emph{intuitive} demonstration of the \emph{physical} unreality of
the gravitational waves (GW's). I have considered there a
spatially limited train $\mathcal{L}$ of running (hypothetical)
GW's -- the source of which is at spatial infinity --, satisfying
exactly the equations $R_{jk}=0$. (It was implicitly assumed that
the $g_{jk}$'s of $\mathcal{L}$ do \emph{not} possess any
singularity of any kind whatever.) Then, the proof rested on a
characteristic property of \emph{general} relativity (GR), that
distinguishes it from Maxwell theory: the absence of any
limitation to the velocities of the reference frames. Thus, we can
ideally consider an observer $\Omega$, who moves together with our
train $\mathcal{L}$. For $\Omega$ the metric tensor of
$\mathcal{L}$ is \emph{time independent}; consequently, Serini's
theorem tells us that its Riemann curvature tensor is zero: the
GW's of $\mathcal{L}$ are mere coordinate undulations.

\par  Of course, this demonstration of the \emph{physical}
non-existence of GW's is a little bold. But there exist absolutely
trenchant proofs, as e.g. the proofs of the non-existence, in the
\emph{\textbf{exact}} formulation of GR, of ``mechanisms'' capable
of generating GW's, \emph{in primis} the \emph{\textbf{fact}} that
the purely gravitational motions of bodies are \emph{geodesic}
\cite{8}. Quite generally, even the \emph{non}-purely
gravitational motions cannot generate GW's, see \cite{9}.

\par A last remark. One could \emph{\textbf{object}} that -- as a
matter of fact -- there are wavy solutions of Einstein equations
$R_{jk}=0$, the curvature tensor of which is different from zero.
\emph{\textbf{Answer}}: \emph{i}) \emph{all} solutions of
$R_{jk}=0$ do not possess an energy-momentum endowed with a
\emph{true} tensor character: accordingly, they are unphysical
objects; \emph{ii}) any undulatory character can be obliterated by
a sequence of suitable coordinate transformations; \emph{iii}) the
\emph{mathematical} existence of wavy solutions of $R_{jk}=0$,
having a curvature tensor $R_{jklm}\neq0$, can be easily
understood: let $W$ be a solution of this kind; it owes its
computative existence to a given gravity source $S$ (explicitly or
implicitly postulated) at a very large distance from an ideal
observer \cite{10}. Of course, $W$ retains ``memory'' of the
spacetime curvature produced by $S$ -- for a detailed and
analytical corroboration of this statement, see e.g. the treatment
given by Fock in Ch. VII of his book \cite{2bis}; on the other
hand, no motion of a gravity source, no cataclysmic disruption of
it can give origin to GW's, as it has been proved.

\vskip0.70cm
\begin{center}
\noindent \emph{\textbf{APPENDIX}} \textbf{B}
\end{center}

The analogy between Maxwell e.m. theory and Einstein general
relativity is a misleading analogy. This is, in particular,
clarified also by the intuitive proof of the non-existence of
physical GW's, which I have recalled in App.A. In his splendid
\emph{Autobiographisches} \cite{11}, at page 53, Einstein
emphasized the following paradox of classical time conception,
which was discovered by him when he was only 16 years old: ``Wenn
ich einem Lichtstrahl nacheile mit der Geschwindigkeit $c$
(Lichtgeschwindigkeit im Vacuum), so sollte ich einen solchen
Lichtstrall als ruhendes, r\"aumlich oszillatorisches
elektromagneti\-sches Feld wahrnehmen. So etwas scheint es aber
nicht zu geben, weder auf Grund der Erfahrung noch gem\"a\ss{} den
Maxwell'schen Gleichungen.'' In the English translation by P.A.
Schilpp: ``If I pursue a beam of light with the velocity $c$
(velocity of light in a vacuum), I should observe such a beam of
light as a spatially oscillatory electromagnetic field at rest.
However, there seems to be no such thing, whether on the basis of
experience or according to Maxwell's equations.'' Now, in
\emph{general} relativity the paradoxical character of the above
consideration \emph{disappears} if in particular the beam of light
is substituted by a beam of (hypothetical) GW's: indeed, in GR
there is no limitation to the velocity of the reference frames.

\end{document}